\documentclass[dvips]{article}
\usepackage{supertabular,lscape,epsfig}
\usepackage{amssymb}
\usepackage{amsmath}

\usepackage[polish]{babel}
\usepackage[T1]{fontenc}
\usepackage[latin2]{inputenc}
\usepackage{pslatex}

\DeclareSymbolFont{ppa}{OT1}{ppl}{m}{it}
\DeclareMathSymbol{\vv}{\mathalpha}{ppa}{'166}

\begin{document}

\newcommand{\dd}{\,{\rm d}}
\newcommand{\ie}{{\it i.e.},\,}
\newcommand{\etal}{{\it et al.\ }}
\newcommand{\eg}{{\it e.g.},\,}
\newcommand{\cf}{{\it cf.\ }}
\newcommand{\vs}{{\it vs.\ }}
\newcommand{\zdot}{\makebox[0pt][l]{.}}
\newcommand{\up}[1]{\ifmmode^{\rm #1}\else$^{\rm #1}$\fi}
\newcommand{\dn}[1]{\ifmmode_{\rm #1}\else$_{\rm #1}$\fi}
\newcommand{\upd}{\up{d}}
\newcommand{\uph}{\up{h}}
\newcommand{\upm}{\up{m}}  
\newcommand{\ups}{\up{s}}
\newcommand{\arcd}{\ifmmode^{\circ}\else$^{\circ}$\fi}
\newcommand{\arcm}{\ifmmode{'}\else$'$\fi}
\newcommand{\arcs}{\ifmmode{''}\else$''$\fi}
\newcommand{\MS}{{\rm M}\ifmmode_{\odot}\else$_{\odot}$\fi}
\newcommand{\RS}{{\rm R}\ifmmode_{\odot}\else$_{\odot}$\fi}
\newcommand{\LS}{{\rm L}\ifmmode_{\odot}\else$_{\odot}$\fi}

\newcommand{\Abstract}[2]{{\footnotesize\begin{center}ABSTRACT\end{center}
\vspace{1mm}\par#1\par   
\noindent
{~}{\it #2}}}

\newcommand{\TabCap}[2]{\begin{center}\parbox[t]{#1}{\begin{center}
  \small {\spaceskip 2pt plus 1pt minus 1pt T a b l e}
  \refstepcounter{table}\thetable \\[2mm]
  \footnotesize #2 \end{center}}\end{center}}

\newcommand{\TableSep}[2]{\begin{table}[p]\vspace{#1}
\TabCap{#2}\end{table}}

\newcommand{\FigCap}[1]{\footnotesize\par\noindent Fig.\  %
  \refstepcounter{figure}\thefigure. #1\par}

\newcommand{\TableFont}{\footnotesize}
\newcommand{\TableFontIt}{\ttit}
\newcommand{\SetTableFont}[1]{\renewcommand{\TableFont}{#1}}

\newcommand{\MakeTable}[4]{\begin{table}[htb]\TabCap{#2}{#3}
  \begin{center} \TableFont \begin{tabular}{#1} #4
  \end{tabular}\end{center}\end{table}}

\newcommand{\MakeTableSep}[4]{\begin{table}[p]\TabCap{#2}{#3}
  \begin{center} \TableFont \begin{tabular}{#1} #4
  \end{tabular}\end{center}\end{table}}

\newenvironment{references}%
{
\footnotesize \frenchspacing
\renewcommand{\thesection}{}
\renewcommand{\in}{{\rm in }}
\renewcommand{\AA}{Astron.\ Astrophys.}
\newcommand{\AAS}{Astron.~Astrophys.~Suppl.~Ser.}
\newcommand{\ApJ}{Astrophys.\ J.}
\newcommand{\ApJS}{Astrophys.\ J.~Suppl.~Ser.}
\newcommand{\ApJL}{Astrophys.\ J.~Letters}
\newcommand{\AJ}{Astron.\ J.}
\newcommand{\IBVS}{IBVS}
\newcommand{\PASP}{P.A.S.P.}
\newcommand{\Acta}{Acta Astron.}
\newcommand{\MNRAS}{MNRAS}
\renewcommand{\and}{{\rm and }}
\section{{\rm REFERENCES}}
\sloppy \hyphenpenalty10000
\begin{list}{}{\leftmargin1cm\listparindent-1cm
\itemindent\listparindent\parsep0pt\itemsep0pt}}%
{\end{list}\vspace{2mm}}
 
\def\TYLDA{~}
\newlength{\DW}
\settowidth{\DW}{0}
\newcommand{\dw}{\hspace{\DW}}

\newcommand{\refitem}[5]{\item[]{#1} #2%
\def\REFARG{#3}\ifx\REFARG\TYLDA\else, {\it#3}\fi
\def\REFARG{#4}\ifx\REFARG\TYLDA\else, {\bf#4}\fi
\def\REFARG{#5}\ifx\REFARG\TYLDA\else, {#5}\fi.}

\newcommand{\Section}[1]{\section{#1}}
\newcommand{\Subsection}[1]{\subsection{#1}}
\newcommand{\Acknow}[1]{\par\vspace{5mm}{\bf Acknowledgements.} #1}
\pagestyle{myheadings}

\newfont{\bb}{ptmbi8t at 12pt}
\newcommand{\xrule}{\rule{0pt}{2.5ex}}  
\newcommand{\xxrule}{\rule[-1.8ex]{0pt}{4.5ex}}  
\def\thefootnote{\fnsymbol{footnote}}
\begin{center}

{\Large\bf
On the Prospects for Detection and Identification\\
\vskip2pt
of Low-Frequency Oscillation Modes in Rotating\\ 
\vskip5pt
B Type Stars}
\vskip1.7cm
{\bf J.~~ D~a~s~z~y~\'n~s~k~a-D~a~s~z~k~i~e~w~i~c~z$^1$,~~W.\,A.~~D~z~i~e~m~b~o~w~s~k~i$^{2,3}$\\
and~~ A.\,A.~~ P~a~m~y~a~t~n~y~k~h$^{3,4}$}
\vskip5mm
{$^1$Instytut Astronomiczny, Uniwersytet Wroc{\l}awski.
ul.~Kopernika~11, Poland\\ 
e-mail: daszynska@astro.uni.wroc.pl\\
$^2$Warsaw University Observatory, Al.~Ujazdowskie~4,
00-478~Warsaw, Poland\\
e-mail:wd@astrouw.edu.pl\\
$^3$ Copernicus Astronomical Center, ul.~Bartycka~18, 00-716~Warsaw,
Poland\\
e-mail: alosza@camk.edu.pl\\
$^4$ Institute of Astronomy, Pyatnitskaya Str.~48, 109017~Moscow, Russia
}
\end{center}

\vspace*{6pt}
\Abstract{We study how rotation affects observable amplitudes of
high-order g- and mixed r/g-modes and examine prospects for their detection
and identification. Our formalism, which is described in some detail,
relies on a nonadiabatic generalization of the traditional
approximation. Numerical results are presented for a number of unstable
modes in a model of SPB star, at rotation rates up to 250~km/s. It is shown
that rotation has a large effect on mode visibility in light and in mean
radial velocity variations. In most cases, fast rotation impairs mode
detectability of g-modes in light variation, as Townsend (2003b) has
already noted, but it helps detection in radial velocity variation. The
mixed modes, which exist only at sufficiently fast rotation, are also more
easily seen in radial velocity. The amplitude ratios and phase differences
are strongly dependent on the aspect, the rotational velocity and on the
mode. The latter dependence is essential for mode identification.}{Stars:
oscillations -- Stars: emission-line, Be -- Stars: rotation}

\Section{Introduction}
Variability with frequencies comparable to rotation frequency has been
found in a number of hot (most often Be) stars. Whether such variability is
caused by slow modes has been debated for some time (see \eg Baade 1982,
Balona 1985). However, recent data from the MOST on $\zeta$~Oph (Walker
\etal 2005a), HD\,163868 (Walker \etal 2005b), and $\beta$~CMi (Saio \etal
2007) revealed rich frequency spectra which may be understood only in terms
of oscillation mode excitation. There are also frequency spectra obtained
from ground-based observations, such as of $\mu$~Eri (Jerzykiewicz \etal
2005), which certainly cannot be explained in terms of a rotational
modulation. A potential for using the abundant frequency data as
constraints on stellar models exists but the prerequisite is mode
identification and it has not been done so far.

Most frequently used method of mode identification employs the data on mode
amplitudes and phases of light variability in various bands and in radial
velocity. Amplitude ratios depend on angular dependence of surface
distortion, which in rotating stars is no longer described by a single
spherical harmonic. The departure becomes large once the angular velocity
of rotation is comparable to oscillation frequency. This is not an unusual
situation for cooler B stars in the main sequence band. Take a model of
6~\MS\ star in the mid of its main sequence evolution. Ignoring effects of
rotation, we find that all dipole modes with period between 1.8~d to 2.8~d
are unstable. Rotation periods in this range correspond to equatorial
velocities between 100~km/s and 150~km/s, which is not high for such a
star.

For low frequency modes the surface dependence may be approximately
described in terms of the Hough functions (see \eg Lee and Saio 1997,
Bildsten \etal 1996, Townsend 2003a). This approximation, called
traditional, was used by Townsend (2003b), who calculated observable
light amplitudes and used the results to address the problem of mode
identification for low frequencies in rotating stars using multicolor
data. It is our experience (\eg Daszy\'nska-Daszkiewicz, Dziembowski and
Pamyatnykh 2005), however, that in the case of B-type pulsators, it is
very important to combine photometric and radial velocity data for a
unique discrimination of excited modes and constraining stellar
parameters. Thus, this work, which may be regarded as an extension of
Townsend's paper, focuses on calculation of radial velocity amplitudes.
We adopt an uniform approach in our calculation of all disk-averaged
amplitudes and it is different from that used by him. Furthermore, in
addition to g-modes, we include r-modes, which become propagative in
stellar radiative envelopes once rotation is fast enough (see Savonije
2005, Townsend 2005b, who uses the term mixed gravity-Rossby modes, Lee
2006).

In Section~2, after specifying assumptions adopted in our calculations, we
summarize formulae for angular dependence of velocity and atmospheric
parameters. Expressions for the light and disk-averaged radial velocity
variations are given in Sections~3 and~4, respectively.  Sections~5, 6 and
7 present numerical results for selected modes in one representative
stellar model considering range of rotation rates but ignoring effects of
changes of centrifugal force on model structure.  Unstable mode properties
are briefly described in Section~5. In Section~6, upon adopting an
arbitrary normalization of linear eigenfunctions, we calculate observable
amplitudes of various modes, which may be excited and detected. Prospects
of mode identification are discussed in Section~7. Examples of diagnostic
diagrams employing amplitude ratios and phase differences are shown there.

\Section{Photospheric Parameter Variations and Velocity Field in the
Traditional Approximation} In our study we adopt the standard
approximations, which are the linear nonadiabatic theory of stellar
oscillation and static plane-parallel atmosphere models. These
approximations are well justified in our application. Modes detected in
slowly pulsating B-type stars have indeed very low amplitudes, the
oscillation periods are much longer than the thermal time scale in the
atmosphere, and vertical variations of mode amplitude are small across
the whole atmosphere. The constant kinematic acceleration is easily
included. Like Townsend (2003b) we adopt the traditional approximation,
which allows to separate latitudinal and radial dependencies of the
pulsational amplitudes. We essentially follow his formalism, expect that
we choose the azimuthal and temporal dependence as $Z=\exp[{\rm
i}(m\varphi-\omega t)]$, which implies $m>0$ for prograde modes and
$m<0$ for retrograde modes.

The displacement vector at the surface may  be written as follows
$${\mbox{\boldmath$\xi$}}(R,\theta,\varphi)=\varepsilon R \left(
\Theta,\frac{\varpi\hat\Theta}{\sin\theta},-{\rm i}
\frac{\varpi\tilde\Theta}{\sin\theta}\right)Z\eqno(1)$$
where $\varepsilon$ is an arbitrary, but small, complex constant, and 
$$\varpi=\frac {GM}{\omega^2R^3}.$$ 
The three functions $\Theta$, $\hat\Theta$, $\tilde\Theta$ describe the
latitudinal dependence of solutions and are the Hough functions, which
are obtained as solutions of the Laplace's tidal equations

$$({\cal D}+ms\mu)\Theta=(s^2\mu^2-1){\hat\Theta},\eqno{\rm (2a)}$$ 
$$({\cal D}-ms\mu){\hat\Theta}=[\lambda(1-\mu^2)-m^2]\Theta,\eqno{\rm (2b)}$$
where $s=2\Omega/\omega$ is called the spin parameter, $\mu=\cos\theta$ and
${\cal D}\equiv(1-\mu^2)\displaystyle\frac{\rm d}{{\rm d}\mu}$. The
equations together with boundary conditions at $\mu=0$ and $\mu=1$ define
the eigenvalue problem on $\lambda$. The third function is given by
$${\tilde\Theta}=-m\Theta+s\mu{\hat\Theta}.\eqno(3)$$
Bildsten \etal (1996), Lee and Saio (1997), and Townsend (2003a)
discussed in great detail the $\lambda(s)$ dependence and asymptotic
properties of the Hough functions. Here, we recall only the essentials.

For specified $m$ and $s\rightarrow0$, there are branches with
$\lambda\rightarrow\ell(\ell+1)$ and $\Theta(\theta)\rightarrow
P_\ell^{|m|}$ (except for normalization). These branches correspond to
g-modes distorted by rotation. For prograde sectorial modes ($m=\ell$),
$\lambda$ slowly decreases with $s$. For all other g-modes, the function
$\lambda(s)$ is increasing quite rapidly. We will identify g-mode
branches by the $\ell$ value at $s=0$. Thus, like in the case of no
rotation, we will use $(\ell,m)$ values as the angular quantum numbers
and refer to $\ell$ as the mode degree. However, now specification of
the angular dependence requires also the value of $s$. The symmetry
about the equatorial plane is determined by the parity of $\ell+|m|$. If
it is even, the functions $\Theta(\theta)$ and $\tilde\Theta(\theta)$
are symmetrical and $\hat\Theta(\theta)$ is antisymmetrical. The
opposite is true if $\ell+|m|$ is odd. The branches for which
$\lambda\rightarrow-\infty$ at $s\rightarrow0$ correspond to r-modes.
For each $m<0$, there is one branch crossing zero at $s=|m|+1$. If
$\lambda>0$, the associated modes become propagatory in the radiative
regions, they may be excited and visible in the light variations.
However, following Lee (2006), we will still call them $r$-modes. The
functions $\Theta(\theta)$ and $\tilde\Theta(\theta)$ are
antisymmetrical and $\hat\Theta(\theta)$ is symmetrical about the
equator for these r-modes.

Upon replacing $\ell(\ell+1)$ with $\lambda(s)$, the nonadiabatic mode
properties may be calculated with a reasonable accuracy using the same
code as for non-rotating stars. This is so because for the mode of our
interest, horizontal flux losses, which are not correctly described, are
small. The latitudinal dependence of perturbed thermodynamical
parameters is then given by $\Theta(\theta)$. Thus, the bolometric flux
perturbation may be written as
$$\frac{\delta {\cal F}_{\rm bol}}{{\cal F}_{\rm bol}}=
\varepsilon f\Theta Z,\eqno(4)$$
where $f$ is a complex quantity determined by solution of linear
nonadiabatic equations and it depends on the stellar model and the mode
parameters $m$, $\lambda$ and $\omega$.

For evaluation of perturbed monochromatic fluxes, we also need the
perturbed gravity, which, as follows from Eq.~(1), is given by
$$\frac{\delta g}{g}=-\varepsilon\left(2+\varpi^{-1}
\right)\Theta Z.\eqno(5)$$
The effect of gravity perturbation plays a relatively small role in
light variability caused by slow modes but it is easy to include. More
important is perturbation of star shape leading to changes in the
projected surface element and the limb-darkening. For both we need the
normal to stellar surface which, as follows from Eq.~(1), is given by
$$\delta{\bf n}_s=-\varepsilon\nabla_H(\Theta Z)=-\varepsilon
\left(0,\frac{\partial\Theta}{\partial\theta},\frac{{\rm
i}m\Theta}{\sin\theta}\right) Z.\eqno(6)$$
The change of the directed element of the surface is
$$\frac{\delta\,{\rm d}{\bf S}}{{\rm d}S}=\varepsilon
\left(2\Theta,-\frac{\partial\Theta}{\partial\theta},-\frac{{\rm
i}m\Theta}{\sin\theta}\right)Z\eqno(7)$$
where ${\rm d}S=R^2\,{\rm d}\mu\,{\rm d}\varphi$.

From Eq.~(1), we also obtain the perturbed pulsation velocity field
as seen from an inertial system ($\varphi_0=\varphi+\Omega t$)
$$\delta{\bf v}=\frac{{\rm d}\mbox{\boldmath$\xi$}}{{\rm d}t}=\left(
\frac{\partial}{\partial t}+\Omega
\frac{\partial}{\partial\varphi_0}\right){\mbox{\boldmath$\xi$}}
=\varepsilon R\left[-{\rm i}\omega\mbox{\boldmath$\xi$}+\Omega
{\rm\bf e}_z\times \mbox{\boldmath$\xi$}\right].\eqno(8)$$
Use of the Lagrangian pulsational velocity is adequate for
representing velocity at the photospheric layer because
${\mbox{\boldmath$\xi$}}$ is nearly constant across the outer
layers for the slow modes considered by us.

\Section{Light Variation}
\subsection{Semi-Analytical Formula}
The most straightforward extension of the expression for the light
variation to the case of rotating stars is through the expansion of the
Hough function into the truncated series of the associated Legendre
functions. This was the way Townsend (2003b) derived his expression. We
write the equivalent expression in the form, which is a straightforward
generalization of our formula (Daszy\'nska-Daszkiewicz, Dziembowski and
Pamyatnykh 2003) derived for the case of modes described by single
spherical harmonic. Now the complex amplitude of the light variation in
the $x$ passband may be expressed as
$${\cal A}_x(i)=\varepsilon\sum_{j=1}^{\infty}
\gamma^m_{\ell_j}(s)Y_{\ell_j}^m (i,0)\left[{\cal D}_{\ell_j}^x
f+{\cal E}_{\ell_j}^x\right]\eqno(9)$$
where
$$\ell_j=\left\{\begin{array}{ll}
 |m|+2(j-1) & {\rm even-parity~modes}\\
 |m|+2(j-1)+1 & {\rm odd-parity~modes}
\end{array}\right.$$
and
$${\cal D}_{\ell}^x=-1.086b_{\ell}^x\frac14\frac{\partial\log
({\cal F}_x|b_{\ell}^x|)}{\partial\log T_{\rm{eff}}},$$
$${\cal E}_{\ell}^x=-1.086b_{\ell}^x\left[(2+\ell)(1-\ell)
-\left(2+\varpi^{-1}\right)\frac{\partial\log({\cal
F}_x|b_{\ell}^x|)}{\partial\log g}\right],$$
$$b_{\ell}^x=\int\limits_0^1 h_x(\tilde\mu)\tilde\mu P_{\ell}(\tilde\mu)
\,{\rm d}\tilde\mu.\eqno(10)$$
where $h_x$ is the limb-darkening law, adopted in the nonlinear form (Claret
2000). With this form Townsend obtained his analytical expressions for
$b_{\ell}^x$. The quantities $\gamma^m_{\ell_j}(s)$, which have to be
calculated numerically, are the expansion coefficients of the $\Theta$
function into the series of the Legendre functions. That is
$$\Theta(\theta)=\sum_{j=1}^{\infty}\gamma_{\ell_j}^m
(s)P_{\ell_j}^m(\theta).\eqno(11)$$
The expression given by Eq.~(9) is quite revealing. At specified
$\varepsilon$, the light amplitudes of low degree modes must decrease with
$s$ because of increasing role of higher order terms which suffer more from
disk-averaging. The higher order terms lead to the aspect-dependence of the
amplitude ratios.

Unfortunately, we could not find a corresponding semi-analytical
expression for the radial velocity and this is why we decided to rely on
two-dimensional numerical integration over the visible hemisphere. We
used Eq.~(9) to check the accuracy.

\subsection{Numerical Approach}
The total flux in the $x$ passband toward the observer is given by
$$L_x=\int\limits_{S}{\cal F}_x h_x{\bf n}_{\rm obs}\cdot{\rm d}{\bf S}\eqno(12)$$
where integration is carried over visible part of star surface, $S$, and
${\bf n}_{\rm obs}$ is the unit vector toward observer.

In the spherical coordinate system with the polar axis parallel to the
rotation axis, we have
$${\bf n}_{\rm obs}\equiv(o_r,o_{\theta},o_{\varphi})\eqno(13)$$
where
$$o_r\equiv\tilde\mu=\cos i\cos\theta+\sin
i\sin\theta\cos(\varphi-\varphi_0),$$
$$o_{\theta}=-\cos i\sin\theta+\sin
i\cos\theta\cos(\varphi-\varphi_0),$$
$$o_{\varphi}=-\sin i\sin(\varphi-\varphi_0).$$
The observer's angular coordinates are $(i,\varphi_0)$.
The first order perturbation of the total flux is given by
$$\delta L_x=\int\limits_S[({\delta\cal F}_x h_x +{\cal F}_x\delta
h_x)\,{\rm d}S{\bf e}_r+{\cal F}_x h_x\delta\,{\rm d}{\bf S}]\cdot{\bf n}_{\rm
obs}.\eqno(14)$$

Assuming an equilibrium atmosphere, we have from Eqs.~(4) and (5)
$${\delta\cal F}_x=\varepsilon\left[\frac{\alpha_T^x}{4}f-\alpha_g^x\left(
2+\varpi^{-1}\right)\right]\Theta Z$$ 
where
$$\alpha_T^x=\frac{\partial\log{\cal F}_x}{\partial\log T_{\rm
eff}}\quad\mbox{and}\quad\alpha_g^x=\frac{\partial\log{\cal
F}_x}{\partial\log g}$$ 
are the derivatives determined numerically from grids of stellar
atmosphere models.
For perturbed limb darkening, we take into account perturbations of the
coefficients on $T_{\rm eff}$ and $g$ though they are only secondary
contributors to light variation. More important contribution arises
from $\delta{\bf n}_s$ (see Eq.~6). With all terms included, we obtain
$$\delta h_x=\varepsilon\left\{\left[\frac{\partial h_x}{\partial\ln
T}\frac{f}{4}-\frac{\partial h_x}{\partial\ln g}\left(2+\varpi^{-1}
\right)\right]-\frac{\partial h_x}{\partial\tilde\mu}\left
(\frac{\partial\Theta}{\partial\theta}o_\theta+\frac{{\rm
i}m\Theta}{\sin\theta}o_\varphi\right)\right\}\Theta Z.$$ 

The derivatives of Claret's $h_x$ are given in Appendix~A1.
\begin{figure}[htb]
\centerline{\includegraphics[width=9cm]{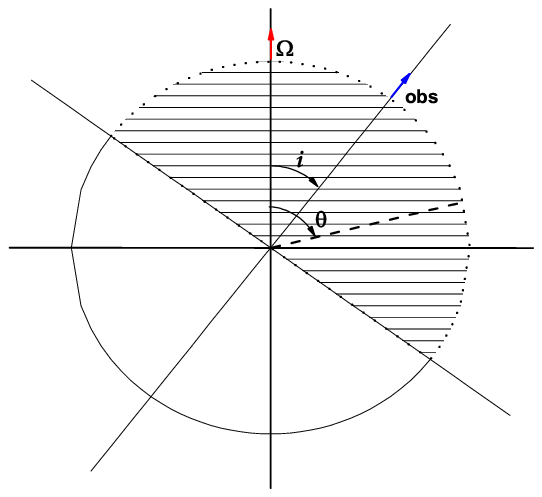}}
\FigCap{The meridional view of the integration area.}
\end{figure}
\begin{figure}[htb]
\centerline{\includegraphics[width=12cm]{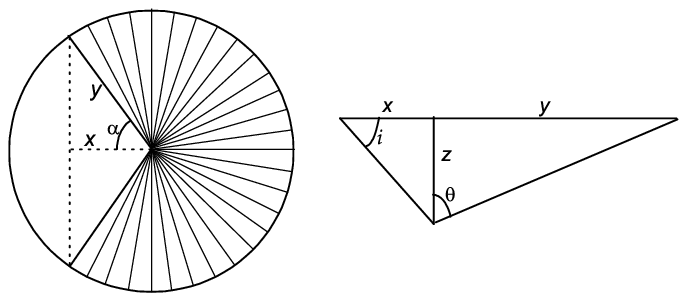}}
\FigCap{Integration area over the azimuthal angle ({\it left}) and the
edge-on view ({\it right}).}
\end{figure}
The expression for perturbed surface element is given in Eq.~(7). The
integration is carried over the unperturbed visible hemisphere. Within the
linear approximation, the integration boundary is unchanged. Note also that
the horizontal component of the displacement does not enter the
expression. The domains of integration over $\theta$ and $\varphi$ are
shown in Fig.~1 and Fig.~2, respectively. From Figs.~1 and 2 it follows
that the ranges are:
$$0\le\theta\le\frac{\pi}2+i,$$ 
$$-(\pi-\alpha)\le\varphi-\varphi_0\le(\pi-\alpha)$$
where $\alpha=\arccos[\cot\theta\cot i]$. For each $\theta$, we
carry the integration over the azimuthal angle,
$\Psi=\varphi{-}\varphi_0$, from $-\beta$ to $+\beta$, where
$\beta=\pi{-}\alpha=\pi{-}\arccos [\cot\theta\cot i]$ (see Fig.~1 and
Fig.~2). We use identities
$$\int\limits_{-\beta}^{\beta}G(\Psi)Zd\Psi=\left\{ 
\begin{array}{ll}
2\int\limits_0^{\beta}G(\Psi)\cos m\Psi\,{\rm d}\Psi & {\rm if}~~G(\Psi)~~{\rm is~even}\\ 
2{\rm i}\int\limits_0^{\beta}G(\Psi)\sin m\Psi\,{\rm d}d\Psi & {\rm if}~~G(\Psi)~~{\rm is~odd}
\end{array}\right.$$
The final expression for the total light variation can be written in
the following form
$$\frac{\delta L_x}{L_x}=\varepsilon\left[\left(
\frac{\alpha_T^x}{4}{\cal B}_1+{\cal B}_3\right)f+2{\cal
B}_1+{\cal B}_2-(2+\varpi^{-1})(\alpha_g^x {\cal B}_1+{\cal
B}_4)\right] Z_0\eqno(15)$$
and $Z_0=\exp[{\rm i}(m\varphi_0-\omega t)]$. In Appendix~A2 we give
explicit expressions for the two-dimensional integrals, ${\cal B}$, which
depend on two angular numbers, spin and the aspect. The integrals take into
account changes in the limb-darkening resulting from the change of the
normal (Eq.~6) as well as the change due to perturbation of the local
temperature (Eq.~4) and gravity (Eq.~5). The two latter are given
through derivatives of $h_x$ with respect to $\log T_{\rm eff}$ and $\log
g$. There are many terms contributing to $\delta L_x$.  However, in our
applications two are far dominant: the one resulting from temperature
perturbation, which is proportional to $\alpha_T^x{\cal B}_1$, and the
other resulting from the surface distortion, which is proportional to
${\cal B}_2$.

\Section{Radial Velocity Variation}
Adopting the standard sign convention, we write the radial
velocity averaged over the stellar disk in the following form:
$$\langle V_{\rm rad}\rangle=-\frac{\int\limits_{S_0}({\bf v}\cdot
{\bf n}_{\rm obs}){\cal F}~h_x~ {\bf n}_{\rm obs}\cdot{\rm d}{\bf S}}
{\int\limits_{S_0}{\cal F}~h_x~{\bf n}_{\rm obs}\cdot{\rm d}{\bf S}}\eqno(16)$$
where for the total velocity field we use
$${\bf v}=\delta{\bf v}+\Omega R\sin\theta{\bf e}_{\varphi}.\eqno(17)$$
The pulsational component, $\delta{\bf v}$, as results from
Eqs.~(1) and (8), is given by
$$\delta{\bf v}=-{\rm i}\omega\varepsilon R\left(
\begin{matrix}
\Theta-\displaystyle\frac{s}2\varpi\tilde\Theta\cr\cr 
\varpi\left[\displaystyle\frac{\hat\Theta}{\sin\theta}
-\displaystyle\frac{s}2\frac{\cos\theta}{\sin\theta}\tilde\Theta\right]\cr\cr 
-{\rm i}\varpi\left[\displaystyle\frac{\tilde\Theta}{\sin\theta}-\displaystyle\frac{s}2
\displaystyle\frac{\cos\theta}{\sin\theta}\hat\Theta\right]+{\rm i}\displaystyle\frac{s}2\Theta\sin\theta
\end{matrix}\right)Z\eqno(18)$$
The contribution of rotation to the mean radial velocity arises from the
same pulsational changes of photospheric parameters which cause luminosity
change and may be calculated in the same way as outlined in
Section~3. Clearly, there is a nonzero contribution only for
non-axisymmetric modes.

We write our final expression for the perturbed radial velocity in the
following form,
$$\delta\langle V_{\rm rad}\rangle={\rm i}\omega\varepsilon
R({\cal C}_{\rm puls}+{\cal C}_{\rm rot})Z_0,\eqno(19)$$ 
with
$${\cal C}_{\rm puls}=\left({\cal B}_5-\frac{s}2{\cal
B}_{7}\right)+\varpi\left({\cal B}_6-\frac{s}2{\cal B}_{8}\right)$$ 
and 
$${\cal C}_{\rm rot}=-\frac{s\sin i}{2}
\left[\left(\frac14\alpha_T^x{\cal B}_9+{\cal B}_{11}\right)f
+2{\cal B}_9+{\cal B}_{10}-\left( 2+\varpi^{-1}\right)
\left(\alpha_g^x{\cal B}_9+{\cal B}_{12}\right)\right].$$
Explicit expressions for the ${\cal B}$ coefficients in terms of the three
Hough functions are given in Appendix~A3.

If $s\approx1$ and $m\neq0$, the two contributions are of the same
order. Similarly to the case of the luminosity, the dominant contributions
to ${\cal C}_{\rm rot}$ are the ones resulting from temperature
perturbation, which here is proportional to $\alpha_T^x{\cal B}_9$, and the
other resulting from the surface distortion, which is proportional to
${\cal B}_{10}$.

\Section{Unstable Modes in a 6~M$_{\pmb\odot}$ Main Sequence Star}
To illustrate how visibility of various modes depends on equatorial
velocity of rotation we choose a model of a 6~\MS\ Population~I star in the
mid of its main sequence evolution. Parameters of the model are given in
Table~1. The model is spherically symmetric, which is consistent with our
use of the traditional approximation, but includes averaged effects of
centrifugal force corresponding to uniform rotation with equatorial
velocity of 250~km/s. The mean effects of centrifugal force are still
reasonably small and therefore we used the same model to study effects of
the Coriolis force at lower equatorial velocities.

\MakeTableSep{crrcrrcccc}{12cm}{Most unstable low degree modes in the B
star model with $M=6.0$~\MS, $\log T_{\rm eff}= 4.205$, $\log
L/\LS=3.204$, for the four values of rotational velocity}
{\hline 
\noalign{\vskip4pt}
$\ell$ & $m$ & $V_{\rm rot}$ & spin & $\lambda~~$ &
$\varpi$ & $f$ & $\nu_{\rm obs}$ [c/d] & $\nu_{\rm star}$ [c/d] &
$\eta~~~$ \\ 
\noalign{\vskip4pt}
\hline
\noalign{\vskip4pt}
  1 & $0$ &   0 &  0.00 &  2.00 &  17.76 & ( 9.99, 11.55) &  0.4220 & 0.4220 & 0.110\\
  1 & $0$ &  50 &  0.86 &  2.33 &  16.35 & (10.56, 12.18) &  0.4396 & 0.4396 & 0.129\\
  1 & $0$ & 150 &  1.97 &  4.62 &   9.48 & ( 8.55, 13.02) &  0.5772 & 0.5772 & 0.202\\
  1 & $0$ & 250 &  2.65 &  7.61 &   6.16 & ( 5.82, 12.46) &  0.7162 & 0.7162 & 0.238\\
\noalign{\vskip4pt}
\hline
\noalign{\vskip4pt}
 1& $+1$ &   0 &  0.00 &  2.00 &  17.76 & ( 9.99, 11.55) &  0.4220 & 0.4220 & 0.110\\
 1& $+1$ &  50 &  1.02 &  1.45 &  22.70 & (10.38, 10.87) &  0.5627 & 0.3731 & 0.063\\
 1& $+1$ & 150 &  3.27 &  1.17 &  26.11 & (10.00, 10.19) &  0.9167 & 0.3478 & 0.031\\
 1& $+1$ & 250 &  5.63 &  1.09 &  27.82 & (10.33, 10.14) &  1.2851 & 0.3369 & 0.021\\
\noalign{\vskip4pt}
\hline
\noalign{\vskip4pt}
 1 & $-1$ &   0 &  0.00 &  2.00 &  17.76 & ( 9.99, 11.55) &  0.4220 & 0.4220 & 0.110\\
 1 & $-1$ &  50 &  0.72 &  3.60 &  11.37 &  (8.98, 12.61) &  0.3375 & 0.5271 & 0.179\\
 1 & $-1$ & 150 &  1.32 & 11.80 &   4.25 &  (3.42, 11.27) &  0.2930 & 0.8619 & 0.258\\
 1 & $-1$ & 250 &  1.71 & 22.21 &   2.57 &  (0.42,  8.50) &  0.0870 & 1.1097 & 0.265\\
\noalign{\vskip4pt}
\hline
\noalign{\vskip4pt}
  2 & $0$ &   0 &  0.00 &  6.00 &   7.35 &  (6.46, 12.38) &  0.6569 & 0.6569 & 0.225\\
  2 & $0$ &  50 &  0.54 &  6.84 &   6.42 &  (5.41, 11.92) &  0.7015 & 0.7015 & 0.234\\
  2 & $0$ & 150 &  1.24 & 13.37 &   3.76 &  (2.43, 10.46) &  0.9170 & 0.9170 & 0.261\\
  2 & $0$ & 250 &  1.66 & 23.56 &   2.42 &  (0.02,  7.93) &  1.1421 & 1.1421 & 0.265\\
\noalign{\vskip4pt}
\hline
\noalign{\vskip4pt}
 2 & $+1$ &   0 &  0.00 &  6.00 &   7.35 &  (6.46, 12.38) &  0.6569 & 0.6569 & 0.225\\
 2 & $+1$ &  50 &  0.58 &  6.02 &   7.28 &  (6.41, 12.36) &  0.8484 & 0.6587 & 0.225\\
 2 & $+1$ & 150 &  1.52 &  8.29 &   5.66 &  (5.11, 12.09) &  1.3162 & 0.7473 & 0.244\\
 2 & $+1$ & 250 &  2.21 & 11.69 &   4.29 &  (3.50, 11.33) &  1.8062 & 0.8580 & 0.257\\
\noalign{\vskip4pt}
\hline
\noalign{\vskip4pt}
 2 & $-1$ &   0 &  0.00 &  6.00 &   7.35 &  (6.46, 12.38) &  0.6569 & 0.6569 & 0.225\\
 2 & $-1$ &  50 &  0.52 &  7.46 &   5.89 & ( 4.74, 11.58) &  0.5426 & 0.7322 & 0.239\\
 2 & $-1$ & 150 &  1.07 & 19.16 &   2.80 & ( 0.68,  8.75) &  0.4937 & 1.0626 & 0.265\\
 2 & $-1$ & 250 &  1.37 & 38.86 &   1.66 & ($-1.79$,4.43) &  0.4315 & 1.3797 & 0.254\\
\noalign{\vskip4pt}
\hline
\noalign{\vskip4pt}
 2 & $+2$ &   0 &  0.00 &  6.00 &   7.35 &  (6.46, 12.38) &  0.6569 & 0.6569 & 0.225\\
 2 & $+2$ &  50 &  0.60 &  5.18 &   7.91 &  (6.23, 11.88) &  1.0110 & 0.6317 & 0.213\\
 2 & $+2$ & 150 &  1.93 &  4.52 &   9.05 &  (7.24, 12.22) &  1.7286 & 0.5908 & 0.202\\
 2 & $+2$ & 250 &  3.29 &  4.31 &   9.49 &  (7.60, 12.32) &  2.4733 & 0.5770 & 0.198\\
\noalign{\vskip4pt}
\hline
\noalign{\vskip4pt}
 2 & $-2$ &   0 &  0.00 &  6.00 &   7.35 &  (6.46, 12.38) &  0.6569 & 0.6569 & 0.225\\
 2 & $-2$ &  50 &  0.51 &  7.57 &   5.81 &  (4.63, 11.51) &  0.3583 & 0.7375 & 0.239\\
 2 & $-2$ & 150 &  1.21 & 14.01 &   3.59 &  (2.08, 10.13) &  0.1992 & 0.9384 & 0.262\\
 2 & $-2$ & 250 &  1.66 & 23.52 &   2.43 &  (0.04,  7.95) &  0.7552 & 1.1411 & 0.265\\
\noalign{\vskip4pt}
\hline
\noalign{\vskip4pt}
 r & $-1$ & 150 &  2.61 &  2.45 &  16.67 & (11.81, 12.77) &  0.1336 & 0.4352 & 0.131\\
 r & $-1$ & 250 &  3.20 &  4.87 &   9.00 & ( 8.12, 12.92) &  0.3557 & 0.5924 & 0.208\\
\noalign{\vskip4pt}
\hline
\noalign{\vskip4pt}
 r & $-2$ & 150 &  3.39 &  1.36 &  28.11 & (14.05, 11.73) &  0.8025 & 0.3352 & 0.040\\
 r & $-2$ & 250 &  3.90 &  3.27 &  13.37 & (11.31, 13.39) &  1.4103 & 0.4859 & 0.164\\
\noalign{\vskip4pt}
\hline}

There are many low frequency modes that are unstable in our selected star
model. These are mainly g-modes. However, at sufficiently high rotation
rate there are also certain r-modes, which become propagatory in radiative
envelope and may become unstable. In all cases, the instability is caused
by the $\kappa$-mechanism acting in the metal opacity bump layer. For
g-modes, the angular order, $\ell$, is defined at the limit of zero
rotation where the angular dependence of amplitude is described by
individual spherical harmonics. The actual mode geometry is determined by
the $(\ell,m)$ numbers and the spin parameter, $s$. Since at each azimuthal
order $m<0$ there is only one r-mode for which $\lambda$ changes sign, we
will identify r-modes by the $m$ value alone. Naturally, we focus attention
on modes suffering least reduction of observable amplitude caused by
disk-averaging. Therefore, we consider g-modes with $\ell\le2$ and r-modes
with $m=-1$ and $-2$.

\begin{figure}[htb]
\centerline{\includegraphics[width=11.5cm]{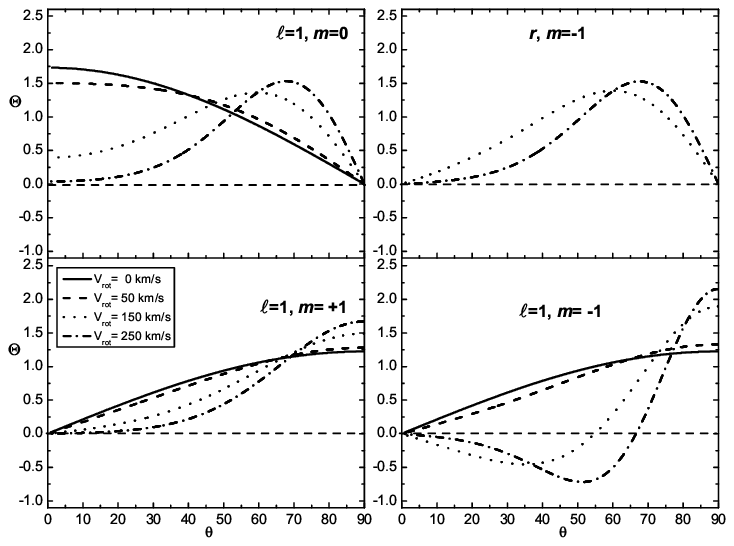}}
\FigCap{The Hough functions for the selected g-modes with
$\ell=1$ and r-modes with $m=-1$ (see Table~1 for more data on the modes).}
\end{figure}
\begin{figure}[htb]
\centerline{\includegraphics[width=11.2cm]{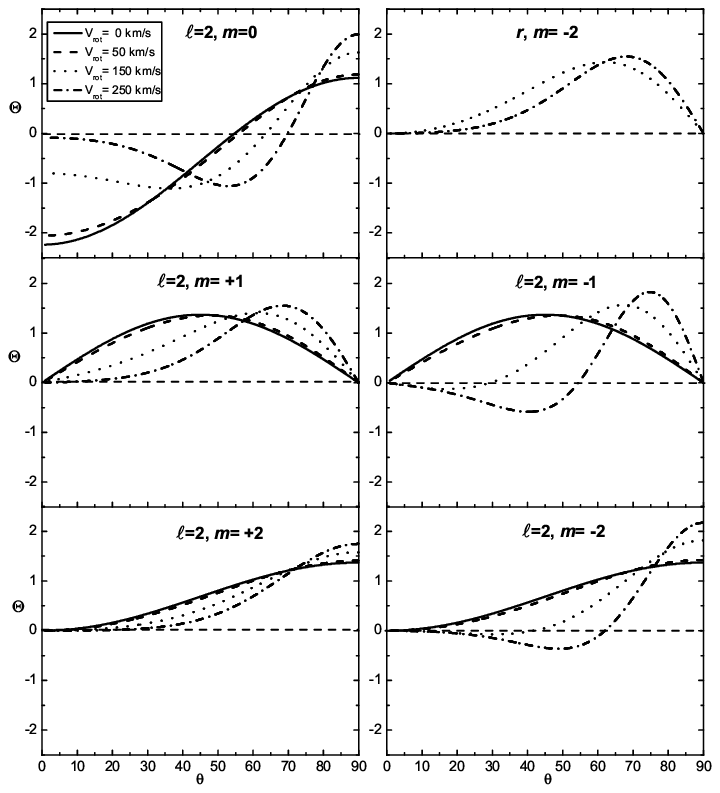}}
\FigCap{Same as Fig.~3 but for the g-modes with $\ell=2$ and
r-modes with $m=-2$.}
\end{figure}
Typically, at each degree and azimuthal order, we find instability
extending over many (up to 40) radial orders. For analysis of visibility,
we selected the mode characterized by the highest normalized growth rate,
$\eta$, which varies between $-1$ and 1. The important parameters of the
selected modes at adopted equatorial velocities are listed in Table~1. For
the r-modes we write "r" instead of $\ell$ as the first entry. The effect
of rotation on mode surface geometry at specified $V_{\rm rot}$ depends on
$s$ which determines $\lambda$. The depth-dependence of eigenfunctions is
determined primarily by the product $\lambda \varpi$. In particular, the
radial orders in this model are given by $n\approx4.23\sqrt{\lambda
\varpi}$. For the selected modes they are between 26 and 34. Nonadiabatic
parameters, $f$ and $\eta$, depend on $n$ but also on the pulsation
frequency in the star reference system, $\nu_{\rm star}$.

The Hough $\Theta$-functions for selected modes are shown in Figs.~3 and
4. We may see the well-known effect of equatorial amplitude confinement,
which increases with $V_{\rm rot}$. The effect is present in all modes
including those with $m=\ell$ though $\lambda(s)$ is a decreasing function
in these cases. As for the remaining Hough functions, which are important,
the confinement is also present. $\tilde\Theta$ and $\Theta$ have the same
symmetry about the equator and the symmetry of $\hat\Theta$ is opposite.
\vspace*{9pt}
\Section{Visibility of Slow Modes}
\vspace*{5pt}
For the modes listed in Table~1, we calculated amplitudes of light
variation in the {\it U} and {\it V} Geneva passbands, $A_U$ and $A_V$,
with Eqs.~(15), as well as that of the radial velocity, $A_{V{\rm rad}}$,
with Eqs.~(19), adopting an arbitrary normalization,
$\varepsilon=0.01$. Coefficients $\alpha_T$ and $\alpha_g$ occurring in
Eq.~(15), were interpolated from the line-blanketed models of stellar
atmospheres (Castelli and Kurucz 2004).

Figs.~5--7 show the $A_V$ and $A_{V{\rm rad}}$ in function of the aspect
angle. Because of the arbitrary normalization, they cannot be regarded as
reliable predictors of expected amplitudes. We may expect that chances of
excitation are related to $\eta$ but it does not determine
$\varepsilon$. Still plots like these provide important hints for
interpretation of the rich oscillation spectra such as that of HD\,163868
(Walker \etal 2005b).
For the $\ell=1$ g-modes we see a simple patterns in the dependence of
$A_V$ on the rotation rate. The aspect-dependence is qualitatively the same
as in the case of no rotation. Rotation gives rise to a departure from the
dipolar angular dependence of $\xi_r$ and $\delta{\cal F}_{\rm bol}$ but
the contribution of the surface distortion to light variation remains small
even at $V_{\rm rot}=250$~km/s. Thus, the $A_V(i)$-dependence is a simple
reflection of the $\Theta(\theta)$-dependence shown in Fig.~3. The
equatorial confinement leads to a reduction of amplitude upon averaging
contributions from the whole hemisphere.
\begin{figure}[htb]
\centerline{\includegraphics[width=11.2cm]{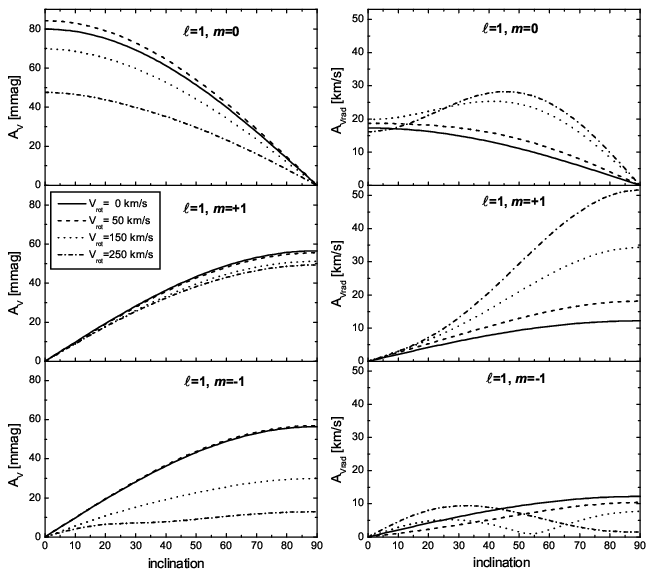}}
\FigCap{Amplitudes of light in the {\it V}-band of Geneva photometry,
$A_V$, and of the radial velocity, $A_{V{\rm rad}}$, at indicated equatorial
rotational velocities plotted as functions of the aspect angle, for the
$\ell=1$ g-modes. The amplitudes are calculated assuming $\epsilon=0.01$
(see Eq.~1).}
\end{figure}

The pattern of the dependence of radial velocity amplitude on the
rotation rate, shown in the right panels of Fig.~5, is more
complicated. In all three cases, the dominant contribution arises from
pulsational velocity (the ${\cal C}_{\rm puls}$ term in Eq.~19). A
secondary but significant contribution, the ${\cal C}_{\rm rot}$ term, adds
in the case of prograde modes and subtracts in the case of retrograde
modes. However, the main reason for the large difference between the modes
and for the nonmonotonic aspect dependence is connected with properties of
${\cal C}_{\rm puls}$. At high rotation rates the contribution from the
advective term in $\delta{\bf v}$ is large and causes that all its three
components play a role. The large values of $A_{V{\rm rad}}$ in the case of
$\ell=1$, $m=+1$ mode for the near equatorial observers arise primarily
from this term. We see that at least for some aspects, fast rotation
increases the chances for mode detection by means of radial velocity
measurements. The $A_V(i)$-dependencies for the $\ell=2$ g-modes depicted
in the left panels of Fig.~6, show that, despite of the increasing
equatorial confinement, the aspect of the best visibility moves toward the
pole with increasing rotation rate. This somewhat unexpected feature, which
has been already noted by Townsend (2003b), is explained in part by the
effect of the surface distortion. There are aspects at which we see
amplitude increase with $V_{\rm rot}$ but on average the effect is the same
as for $\ell=1$, that is fast rotation decreases chances for photometric
detection of slow modes. Also on average, amplitudes of the $\ell=2$ modes
are lower than those of $\ell=1$. In the right panels of Fig.~6, we may see
that radial velocity amplitudes of prograde modes increase with the
rotation rate and that the effect is opposite for the retrograde modes. The
role of the ${\cal C}_{\rm rot}$ term is much more significant than at
$\ell=1$.
\begin{figure}[h]
\centerline{\includegraphics[width=11.3cm]{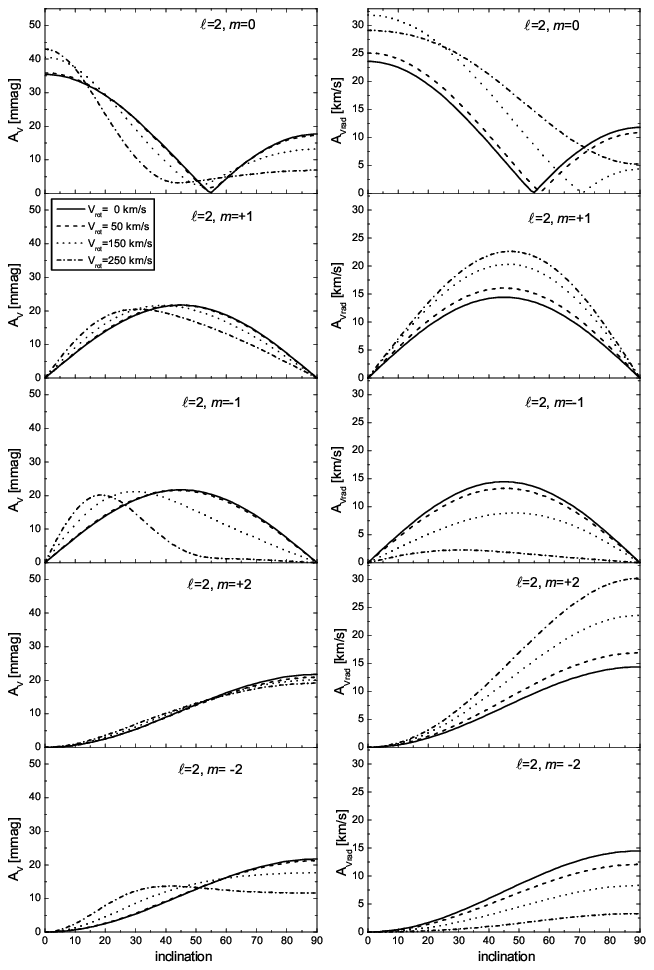}}
\FigCap{Same as Fig.~5 but for the $\ell=2$ g-modes.}
\end{figure}

\newpage
\begin{figure}[htb]
\centerline{\includegraphics[width=11.5cm]{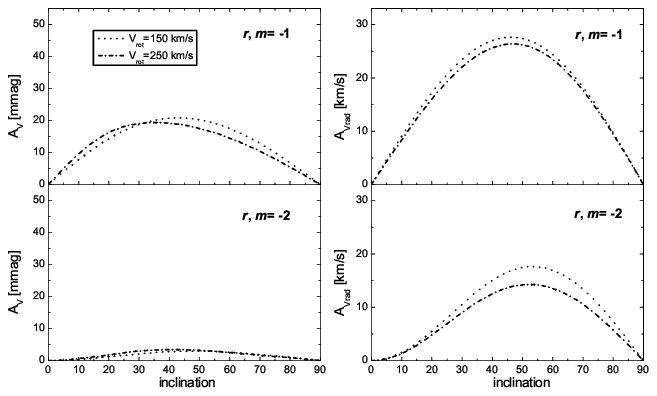}}
\FigCap{Same as Fig.~5 but for the r-modes with $|m|\le2$.}
\end{figure}

We may see in Fig.~3 that $\Theta(\theta)$ for the $m=-1$ r-mode and for
the $\ell=1,~m=0$ g-mode look very similar. Likewise, in Fig.~4 we see the
similarity of $\Theta$'s for the $m=-2$ r-mode and for the $\ell=2, m=+1$
g-mode. Yet, as we may see in Fig.~7, the photometric amplitudes of the
r-modes are much smaller and have different aspect dependence. The
amplitude reduction is in part caused by cancellations in the integral over
$\phi$. Moreover, the effect of distortion, which is significant, cancels
a part of the effect of temperature perturbation. The r-modes are
antisymmetric with respect to the equator and are best seen from the
intermediate aspect angles. The radial velocity amplitudes, shown in the
right panels of Fig.~7, are much less reduced. Thus, spectroscopy gives a
better chance for detecting r-modes.
\vspace*{9pt}
\Section{Prospects for Mode Discrimination}
\vspace*{5pt}
Rotation has a very profound effect on slow mode visibility and hence on
the procedure of mode identification. Unlike in non-rotating stars, the
amplitude ratios depend on the azimuthal order, $m$, and on the aspect,
$i$. Modes with various $m$ differ not only in the surface geometry but
also in their nonadiabatic properties. There are constraints on $m$
following from localization in the frequency spectrum. Modes differing in
the sign of $m$ are well separated. As we have seen in Figs.~5--7, the
aspect is an important factor in mode selection which has to be taken into
account considering possible $m$ and $i$ values. Nonetheless, a unique
discrimination of modes is not likely possible without employing amplitude
and phase data.

Let us begin with photometric data alone. The observables, which do not
depend on $\varepsilon$, are amplitude ratios and phase differences from
multiband photometry. In this case, the potential for mode discrimination
rests mainly on the difference in the relative temperature and distortion
contribution to the total light variation in different passbands. The
latter contribution for the $\ell=1$ g-modes is small, hence a
discrimination between such modes might be difficult. However, as the plots
in Fig.~8 show, these modes may be rather easily distinguished from the
$\ell=2$ and r-modes even if information about $i$ and $V_{\rm rot}$ is
imprecise. A discrimination between the two $\ell=2$ modes could be
difficult if the measured phase difference is between 0.2 and 0.3. The
range is consistent with $m=-1$, if $V_{\rm rot}$ is between 150~km/s and
175~km/s, and $m=-2$, if $V_{\rm rot}$ is 250~km/s.
\begin{figure}[htb]
\centerline{\includegraphics[width=12.3cm]{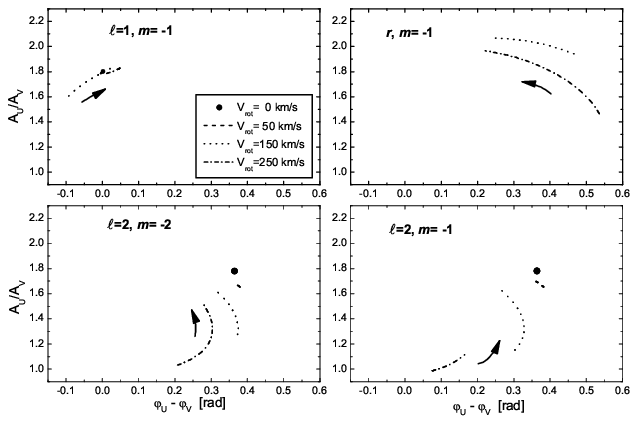}}
\FigCap{Photometric diagnostic diagrams for selected modes based
on light amplitudes and phases in the {\it U} and {\it V} Geneva passbands,
for retrograde g-modes and for the r-mode with $m=-1$. The arrows indicate
direction of increasing $i$. The $i$-ranges are limited by the condition
$A_V>10$~mmag}
\end{figure}

Any discrimination between these four modes based on frequencies alone
would not be possible because all of them are retrograde and occupy the low
frequency end of the oscillation spectrum. It is important to remember that
the plots refer only to most unstable modes of each type. Thus, in real
life, there is additional uncertainties in frequencies and in the
$f$-values that affect the amplitude ratios and phase differences.
With mean radial velocity data, discrimination between the two $\ell=2$
modes should be unambiguous, as the plots in the lower panels of Fig.~9
show. In the upper panels we see that also distinguishing the two $\ell=1$
modes should be easy. Disentangling the $m=+1$ and $m=0$ cases,
however, may require data on line profile changes.
\begin{figure}[htb]
\centerline{\includegraphics[width=12cm]{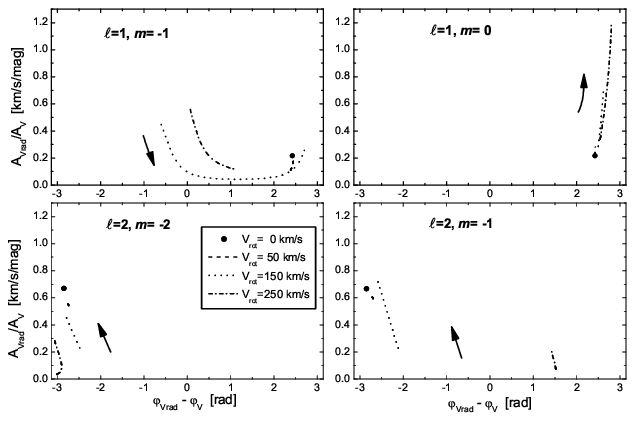}}
\FigCap{Diagnostic diagrams for selected modes based on amplitudes
and phases in the {\it V} Geneva photometric passband and in the radial
velocity, for g-modes with $\ell=1$ and~2.}
\end{figure}

The main benefit from radial velocity data is the possibility of
simultaneous determination of $\ell$ with no need for specifying the value
of the complex parameter $f$. Instead, as described in the case of
non-rotating stars by Daszy\'nska-Daszkiewicz \etal (2005), the combined
spectroscopy and photometry data on amplitudes and phases may be used to
determine $f$, which becomes an independent seismic observable, and the
mode degree, $\ell$. In order to see how the method may be extended to the
present case, let us rewrite Eqs.~(15) and (19) as expressions for the
complex amplitudes of light in the $x$-band and of radial velocity,
respectively. From Eq.~(15) we obtain
$${\cal A}_x(i)={\cal D}^x(i)f\tilde\varepsilon+{\cal
E}^x(i)\tilde\varepsilon\eqno(20)$$ 
where 
$${\cal D}^x(i)=-1.086\left(\frac{\alpha_T^x}{4}{\cal B}_1 +{\cal
B}_3\right),$$
$${\cal E}^x(i)=-1.086[2{\cal B}_1+{\cal B}_2 -(
2+\varpi^{-1})(\alpha_g^x {\cal B}_1+{\cal B}_4)]$$ 
and $\tilde\varepsilon=\varepsilon\,{\rm exp}({\rm i}m\varphi_0)$.
Similarly, from Eq.~(19) the first moment of spectral line is equal to
$${\cal M}_1^x(i)={\rm i}\omega R[{\cal H}^x(i)f\tilde\varepsilon+
{\cal G}^x(i)\tilde\varepsilon]\eqno(21)$$
where 
$${\cal H}^x(i)=-\frac{s \sin i}{2}\left(\frac14\alpha_T^x
{\cal B}_9 +{\cal B}_{11}\right)$$ 
and 
$${\cal G}^x(i)={\cal B}_5{-}\frac{s}2{\cal B}_{7}+\varpi \left({\cal
B}_6{-}\frac{s}2{\cal B}_{8}\right){-}\frac{s\sin i}{2}\left[2{\cal B}_9+{\cal
B}_{10}{-}\left(2+\varpi^{-1}\right)\left(\alpha_g^x{\cal B}_9+{\cal
B}_{12}\right)\right].$$ 
These two equations are counterparts of Eqs.~(1) and (2) of
Daszy\'nska-Daszkiewicz \etal (2005).

In the case of negligible rotation, the dependence of amplitudes on $m$ and
$i$ has been absorbed in $\tilde\varepsilon$ and the problem was reduced to
finding $\tilde\varepsilon f$ and $\tilde\varepsilon$ by the least square
minimization of $\chi^2(\ell)$. The method requires stellar atmosphere
parameters, which in fact may be improved on the process of pulsation
amplitude fitting. An unique determination of $\ell$ is possible even with
imprecise atmosphere parameters if the $\min(\chi^2)(\ell)$ dependence is
strong. In the present case we have two more stellar parameters to improve
which are $i$ and $V_{\rm rot}$. Moreover, we have to take into account the
$m$-dependence. Prospects for mode identification depend on the strength of
the the $\min(\chi^2)(\ell,m)$ dependence. The plots in Figs.~8 and 9
suggest that it is likely the case. However, it remains to be seen when the
method is applied to real data.

\Section{Summary and Conclusions}
Our goal was to examine chances for detection and identification of slow
oscillation modes whose frequencies are of the order of angular velocity of
rotation. In our calculations of expected mode amplitudes, we relied on a
nonadiabatic generalization of the traditional approximation, similar to
that introduced by Townsend (2005a). The chances for detecting a particular
mode depend, in part, on its intrinsic amplitudes, which may be calculated
only in the framework of a nonlinear modeling. Our linear models give us
only a hint which is the growth rate. Such models are expected to be
adequate for describing geometry of the mode which has important impact on
mode visibility, its aspect and the angular degree dependence. With the
formalism outlined in Sections 3 and 4, we calculated the observable
amplitudes for selected unstable modes in models of a 6~\MS\ main sequence
star as a function of the rotation rate and the aspect angle. The model may
be regarded as representative for SPB variables.

Departure from the individual spherical-harmonic dependence which increases
with the rotation rate leads, in most cases, to lower photometric
amplitudes. In contrast, the effect on radial velocity amplitude is most
often opposite. However, the light to radial velocity amplitude ratio
changes significantly from mode to mode and depends on the aspect. We
showed, in particular, that the mixed r/g-modes are most easily detectable
in radial velocity. Considering possible identification of peaks in rich
oscillation spectra such as of HD\,163868 (Walker \etal 2005b) it is
important to take into account the aspect dependence of considered mode
amplitudes as it has been done by Dziembowski \etal (2007).

The observables yielding numerical constraints on mode geometry and the
aspect angle are amplitude ratios and phase differences. We calculated
examples of diagrams based on photometric data in two passbands and on mean
radial velocity measurements. Not always photometric data are sufficient
for mode discrimination. Radial velocity data not only help discrimination
but also allow to proceed in a less model-dependent manner. The
photospheric value of the complex radial eigenfunction corresponding to the
radiative flux may then be determined from data rather than taken from
linear nonadiabatic calculations. A comparison of calculated and deduced
values yields a constraint on the model.

\Acknow{This work has been supported by the Polish MNiSW grant 
No 1 P03D 021 28.}

\newpage
\centerline{\bf Appendix}
\vskip6pt
{\it A1. The Limb-Darkening Law}
\vskip6pt
For calculating the surface integrals in Eq.~(15) and Eq.~(19), we need to
specify the limb-darkening law. Here, we use the nonlinear Claret (2000)
formulae which we rewrite in the following form
$$h_x(\tilde\mu)=2~\frac{1-\sum\limits_{k=1}^4a_k^x(1
-\tilde\mu^{k/2})}{1-\sum\limits_{k=1}^4\displaystyle\frac{k}{k+4}a_k^x }.$$
The derivative with respect to $\tilde\mu$ can be easily obtained from this
formula. The derivatives with respect to the effective temperature and
gravity are given by:
$$\frac{\partial h_x}{\partial\ln T_{\rm eff}}=\frac1{1-
\sum\limits_{k=1}^4\displaystyle\frac{k}{k+4}a_k^x}\cdot\sum_{k=1}^4\left[
\frac{k}{k+4}h_x-2(1-\tilde\mu^{k/2})\right]\frac{\partial
a_k^x}{\partial\ln T_{\rm eff}},$$
and
$$\frac{\partial h_x}{\partial\ln g}=\frac1{1-
\sum\limits_{k=1}^4\displaystyle\frac{k}{k+4}a_k^x}\cdot\sum_{k=1}^4\left[
\frac{k}{k+4}h_x-2(1-\tilde\mu^{k/2})\right]\frac{\partial
a_k^x}{\partial\ln g}$$
respectively.

\vspace{0.5cm}
{\it A2. Integrals in the Expression for the Light Variation}

$${\cal B}_1=\int\limits_0^{\frac{\pi}2+i}\Theta{\cal P}_1\sin\theta\,{\rm d}\theta$$
$${\cal B}_2=\int\limits_0^{\frac{\pi}2+i}\left(\frac{\partial\Theta}{\partial\theta}\sin\theta{\cal P}_2+m\Theta{\cal P}_3\right){\rm d}\theta$$
$${\cal B}_3=\int\limits_0^{\frac{\pi}2+i}\Theta{\cal P}_4\sin\theta\,{\rm d}\theta$$
$${\cal B}_4=\int\limits_0^{\frac{\pi}2+i} \Theta {\cal P}_5\sin\theta\,{\rm d}\theta$$
where
$${\cal P}_1=\frac1{\pi}\int\limits_0^{\beta}\cos m\Psi h_xo_r\,{\rm d}\Psi$$
$${\cal P}_2=-\frac1{\pi}\int\limits_0^{\beta}\cos m\Psi\left(o_r\frac{{\rm d}h_x}{{\rm d}o_r}+h_x\right)o_{\theta}\,{\rm d}\Psi$$
$${\cal P}_3=\frac1{\pi}\int\limits_0^{\beta}\sin m\Psi\left(o_r\frac{{\rm d}h_x}{{\rm d}o_r} +h_x\right) o_{\varphi}\,{\rm d}\Psi$$
$${\cal P}_4=\frac1{\pi}\int\limits_0^{\beta} \cos m\Psi \frac{\partial h_x}{\partial\ln T_{\rm eff}}o_r\,{\rm d}\Psi$$
$${\cal P}_5=\frac1{\pi}\int\limits_0^{\beta} \cos m\Psi \frac{\partial h_x}{\partial\ln g}o_r\,{\rm d}\Psi.$$
The function $\Theta$ is one of three Hough functions (see Section~2). The
components of the unit vector directed to the observer ($o_r, o_{\theta},
o_{\varphi)}$) are given in Eq.~(13).

\vspace{0.5cm}
{\it A3. Integrals in the Expression for the Radial Velocity Variation}

$${\cal B}_5=\int\limits_0^{\frac{\pi}2+i}\Theta {\cal P}_6\sin\theta \,{\rm d}\theta$$
$${\cal B}_6=\int\limits_0^{\frac{\pi}2+i}({\hat\Theta} {\cal P}_7+{\tilde\Theta}{\cal P}_8)\,{\rm d}\theta$$
$${\cal B}_7=\int\limits_0^{\frac{\pi}2+i}\Theta{\cal P}_8\sin^2\theta\,{\rm d}\theta$$
$${\cal B}_8=\int\limits_0^{\frac{\pi}2+i}\left[{\tilde\Theta}{\cal P}_6\sin\theta+\left(\tilde\Theta{\cal P}_7+\hat\Theta{\cal P}_8\right)\cos\theta\right]\,{\rm d}\theta$$
$${\cal B}_9=\int\limits_0^{\frac{\pi}2+i}\Theta{\cal P}_9\sin^2\theta \,{\rm d}\theta$$
$${\cal B}_{10}=\int\limits_0^{\frac{\pi}2+i}\left(\frac{\partial\Theta}{\partial\theta}\sin\theta{\cal P}_{10}+m\Theta{\cal P}_{11}\right)\sin\theta\,{\rm d}\theta$$
$${\cal B}_{11}=\int\limits_0^{\frac{\pi}2+i}\Theta{\cal P}_{12}\sin^2\theta\,{\rm d}\theta$$
$${\cal B}_{12}=\int\limits_0^{\frac{\pi}2+i}\Theta{\cal P}_{13}\sin^2\theta\,{\rm d}\theta$$
where
$${\cal P}_6=\frac1{\pi}\int\limits_0^{\beta}\cos m\Psi h_xo_r^2\,{\rm d}\Psi$$
$${\cal P}_7=\frac1{\pi}\int\limits_0^{\beta}\cos m\Psi h_xo_ro_{\theta}\,{\rm d}\Psi$$
$${\cal P}_8=\frac1{\pi}\int\limits_0^{\beta}\sin m\Psi h_xo_ro_{\varphi}\,{\rm d}\Psi$$
$${\cal P}_9=\frac1{\pi}\int\limits_0^{\beta}\sin m\Psi\sin\Psi h_xo_r\,{\rm d}\Psi$$
$${\cal P}_{10}=-\frac1{\pi}\int\limits_0^{\beta}\sin m\Psi\sin\Psi\left(o_r\frac{{\rm d}h_x}{{\rm d}o_r}+h_x\right)o_{\theta}\,{\rm d}\Psi$$
$${\cal P}_{11}=-\frac1{\pi}\int\limits_0^{\beta}\cos m\Psi\sin\Psi\left(o_r\frac{{\rm d}h_x}{{\rm d}o_r}+h_x\right)o_{\varphi}\,{\rm d}\Psi$$
$${\cal P}_{12}=\frac1{\pi}\int\limits_0^{\beta}\sin m\Psi\sin\Psi\frac{\partial h_x}{\partial\ln T_{\rm eff}}o_r\,{\rm d}\Psi$$
$${\cal P}_{13}=\frac1{\pi}\int\limits_0^{\beta}\sin m\Psi\sin\Psi\frac{\partial h_x}{\partial\ln g}o_r\,{\rm d}\Psi$$
\end{document}